\documentclass[onecolumn,12pt,preprintnumbers,aps]{revtex4}
\begin{document}
\title{Evidence of Ferrimagnetism in Ferromagnetic La$_{0.67}$Ca$_{0.33}$MnO$_{3}$ nanoparticle}
\author{R.N. Bhowmik\footnote{E-mail address for correspondence:\\rnbhowmik.phy@pondiuni.edu.in}}

\address{Department of Physics,Pondicherry University, R.V.
Nagar, Kalapet, Pondicherry-605014, India}

\begin{abstract}
The present report is dedicated to show that ferromagnetic
La$_{0.67}$Ca$_{0.33}$MnO$_{3}$ (LCMN) particles can be better
described in the framework of ferrimagnetic model. To confirm the
ferrimagnetic signature in ferromagnetic LCMN particles, the
temperature dependence of the inverse of magnetic susceptibility in
the paramagnetic state of the samples was taken as a tool of data
analysis. The observed ferrimagnetism is understood as an effect of
of the core-shell spin structure in LCMN particles.\\

Key Words: A. Ferromagnetic nanoparticle; B. Mechanical Milling; C.
Ferrimagnetism; D. Core-shell spin structure\\

PACS:75.47.Lx,75.30.Cr,75.50.-y,75.50.Gg,75.50.Tt
\end{abstract}
\maketitle

\section{Introduction}
Magnetic nanomaterials are continued to be at the center of current
research interests due to their huge technological applications and
incomplete undestanding of many discovered phenomena. For example,
superparamagnetic blocking of magnetic moments below the
conventional paramagnetic to ferromagnetic transition temperature
(T$_C$), appearance of unconventional spin glass behaviour at lower
temperatures, decrease of effective magnetic moment of the material,
exchange bias effect, quantum tunnelling of magnetization, and low
field magnetoresistance have been observed when the particle size of
ferromagnetic materials decreases into nanosize dimension (below 100
nm)\cite{Ramirez,Gubin}. Various mechanisms have been introduced in
literature to describe the magnetic features of nanoparticles, e.g.,
core-shell structure, dipole interactions, inter-particle
interactions, exchange anisotropy \cite{Battle}. Among the proposed
mechanisms, the core-shell concept is world wide accepted to explain
the features of nanoparticle magnetism. In a magnetic nanoparticle
the central part, known as core, is assumed to be identical to the
structure and property of bulk material with micron sized particles.
The structure and property of the outer part of the particle, known
as shell, are drastically different in comparison with core
\cite{Battle,CoRh2O4}. If the bulk material is a typical long ranged
ferromagnet (antiferromagnet), then core is assumed to be long
ranged ferromagnet(antiferromagnet) and disorder is introduced in
the shell part of the particle. This means the property of a
magnetic nanoparticle is basically heterogeneous in character (i.e.,
consisting of two different magnetic components or equivalent to two
magnetic sublattices) over a length of particle dimension and also
in the whole dimension of the material when the particles are in
contact. The common phenomena due to the heterogeneous magnetic
structure in ferromagnetic nanoparticles are the reduction of
particle moment and magnetic blocking/freezing at lower
temperatures. On the other hand, antiferromagnetic nanoparticles
have shown many enhanced properties mainly due to different magnetic
structure of shell part in comparison with bulk counter material
\cite{CoRh2O4}. This shows that core-shell structure plays an
important role in the properties of magnetic materials, immaterial
of ferromagnetic or antiferromagnetic particles. Hence, proper
understanding of the effects of core-shell structure is not only the
long standing problem, but also useful in designing the application
oriented materials. To understand the effects of core-shell
structure in different types heterogeneous magnetic structures,
e.g., ferromagnetic core is surrounded by
antiferromagnetic/paramagnetic/ferrimagnetic shell or
antiferromagnetic core is surrounded by ferromagnetic/ferrimagnetic
shell have been synthesized and reported in literature
\cite{Bahl,Vasil,Muri,Masala}. The effect of shell disorder and spin
frustration has also been discussed in many spin-bilayer magnetic
systems \cite{Battle}. G. Bouzerar et al. \cite{Bouz} discussed the
effect of competition between introduced superexchange
(antiferromagnetic) interactions in long ranged double exchange
ferromagnetic matrix. They argued that in the lower limit of
anitiferromagnetic superexchange interactions the long ranged
ferromagnetic state is not altered significantly; rather a canted
ferromagnetic phase or induced new magnetic phase is appeared in the
spin system. The induced magnetic phases may be either stable or
unstable depending on the quantum of magnetic disorder and
frustration. Some report also studied core-shell structure in a
composite material consisting of ferrimagnetic core and
ferroelectric shell
\cite{Coral}.\\
Recently, La$_{0.67}$Ca$_{0.33}$MnO$_3$ nanoparticles in crystalline
and amorphous structural phases have shown many interesting magnetic
properties, related to magnetic disorder at core-shell structure of
the particles \cite{rnbJAP}. A proper knowledge of magnetic
interactions between core-shell spins would be useful not only to
realize the colossal magnetoresistance and inter-grain tunneling of
polarized spins, but also relevant to realize the effect of disorder
on double exchange ferromagnetism in manganites. In the present
work, we demonstrate that the modified magnetism in ferromagnetic
La$_{0.67}$Ca$_{0.33}$MnO$_3$ nanoparticles is identical to the
typical features of ferrimagnetic materials. The evidence of
ferrimagnetic signature in La$_{0.67}$Ca$_{0.33}$MnO$_3$
nanoparticles is also discussed by comparing the features already
observed in ferrimagnetic (Mn$_{0.5}$Ru$_{0.5}$Co$_2$O$_4$ and
MnCr$_2$O$_4$) particles.

\section{Experimental}
Details of the sample preparation of La$_{0.67}$Ca$_{0.33}$MnO$_3$
(perovskite) particles and their characterization have been reported
elsewhere \cite{rnbJAP}. In brief, the polycrystalline bulk
La$_{0.67}$Ca$_{0.33}$MnO$_3$ sample was prepared by solid state
sintering (maximum temperature 1380$^0$C) method. The bulk sample
was subjected to mechanical milling in Argon atmosphere upto 200
hours using Fritsch Planetary Mono Mill "Pulverisette 6" to
synthesize the material in nanocrystalline and amorphous phase. The
structural phase of the samples was confirmed from room temperature
XRD spectrum. The XRD spectrum indicated that crystalline nature of
the material decreases significantly for the milling time more than
61 hours and amorphous phase dominates in the spectrum for milling
time more than 98 hours. Both bulk and milled samples (upto mh98)
are in similar crystallographic phase and found to be matching with
orthorhombic structure with Pnma space group. The temperature
dependence of magnetization under zero field cooled condition was
measured using SQUID magnetometer (MPMS-Quantum Design, USA). The
temperature dependence of dc magnetization at 100 Oe in the
temperature range 100 K to 400 K was also reproduced using vibrating
sample magnetometer (Lakeshore 7404 model).

\section{EXPERIMENTAL RESULTS}
Details of the temperature and field dependence of dc magnetization
have been reported elsewhere \cite{rnbJAP}. In summary, the
paramagnetic to ferromagnetic Curie temperature (T$_C$) for bulk
(LCMN) sample is nearly 281 K and T$_C$ decreases to 262 K, 250 K,
238 K, 225 K and 212 K for mechanical milled mh25(nanocrystalline,
particle size $\sim$ 65 nm), mh61 (nanocrystalline, particle size
$\sim$ 12 nm), mh98 (nanocrystalline+amorphous, particle size $\sim$
16 nm), mh146 (amorphous, particle size $\sim$ 60 nm) and mh200
(amorphous, particle size $\sim$ 90 nm) samples. At the same time,
the long ranged ferromagnetic order (spontaneous magnetization
$\sim$ 3.6 ${\mu}_B$) of bulk LCMN sample decreases to 2.17, 0.87,
0.35, 0.17, 0.10 (in ${\mu}_B$) unit) for mh25, mh61, mh98, mh146
and mh200 samples, respectively. These are some typical features of
the magnetic disorder effect in ferromagnetic materials. In the
present paper, we would like to show some specific magnetic features
of the samples based on magnetization data at (higher temperature)
paramagnetic regime. In Fig. 1, the dc magnetic susceptibility
($\chi_{dc}$ = M/H) of bulk LCMN sample sharply increases above the
magnetization peak temperature T$_p$ $\sim$ 260 K. On the other
hand, magnitude of susceptibility, as well as sharp increase of
magnetization below the respective T$_C$ systematically decreases
for mh25, mh98 and mh200 samples. The decrease of the $\chi$ (T)
variation in milled samples reflects the increasing magnetic
disorder in the ferromagnetic material and realized in the previous
work \cite{rnbJAP}. Interestingly, a typical ferrimagnetic sample,
e.g., Mn$_{0.5}$Ru$_{0.5}$Co$_2$O$_4$ (RuMn) spinel oxide in the
inset of Fig. 1, also exhibits the similar $\chi_{dc}$ (T) behaviour
above its magnetization peak temperature T$_p$ $\sim$ 100 K. This
means only the shape of $\chi_{dc}$ (T) curve in the paramagnetic
state can not determine the nature of magnetic order in the samples,
whether ferromagnet or ferrimagnet. The nature of magnetic order can
be confirmed in convincing manner from the temperature dependence of
the inverse of magnetic susceptibility in the paramagnetic state.
For this purpose, we extended the dc magnetization measurement up to
400 K. First, we confirm the difference of the temperature
dependence of the inverse of susceptibility curve in paramagnetic
regime between bulk LCMN (ferromagnetic) and RuMn (ferrimagnetic)
samples. Fig. 2 shows that the inverse of dc susceptibility
($\chi^{-1}_{dc}$ = H/M) data for bulk LCMN sample at high
temperatures (T $\geq$ 300 K) are fitted with a simple Curie-Weiss
law:
\begin{equation}
 \chi = C/(T-\theta_w)
\end{equation}
Application of this equation confirms the ferromagnetic order in
bulk LCMN sample. The obtained parameters are Curie constant (C
$\sim$ 0.0196) and paramagnetic Curie temperature ($\theta_w$ $\sim$
+ 270 K). In contrast, the inverse of dc susceptibility
($\chi^{-1}$) for Mn$_{0.5}$Ru$_{0.5}$Co$_2$O$_4$ spinel oxide at
high temperatures is fitted with a typical equation:
\begin{equation}
1/\chi = (T-\theta_{1})/C_{eff} - \xi/(T-\theta_{2})
\end{equation}
In general, this equation is applicable for ferrimagnet
\cite{ferri}. The obtained parameters ($\theta_{1}$ $\sim$ -1320 K,
C$_{eff}$ $\sim$ 0.076, $\xi$ $\sim$ 200850, $\theta_{2}$ $\sim$ +
112 K), in particular the positive value of $\theta_{2}$ (slightly
larger than T$_C$ $\sim$ 100 K) and a high negative value of
$\theta_{1}$, clearly indicate the ferrimagnetic order in
Mn$_{0.5}$Ru$_{0.5}$Co$_2$O$_4$ spinel oxide. Similar
($\chi^{-1}_{dc}$ (T))character was also noted in many other
ferrimagnetic materials \cite{ferri,rnbMnCr2O4}.\\
Now, we analyze the temperature dependence of the inverse of dc
susceptibility data for mechanical millled nanoparticle samples. As
shown in Fig. 3, the data are well fitted with a simple Curie-Weiss
law (equation (1)) above 330 K. The fit parameters of Curie-Weiss
law (C and $\theta_w$) are shown in Table I. On the other hand, the
hyperbolic shape of $\chi^{-1}$ (T) curves (with down curvature)
above the Curie temperature of the samples suggests that milled
samples belong to the class of either ferrimagnet or double exchange
ferromagnet \cite{Anderson}. We noted that the $\chi^{-1}$ (T)
curves of the present nanoparticle samples are identical to the
ferromagnetic MnCr$_2$O$_4$ nanoparticle samples \cite{rnbMnCr2O4}.
To clarify the ferrimagnetic nature of the nanoparticle (NP)
samples, we have fitted the $\chi^{-1}$ (T) data in the temperature
range 330 K-400 K using equation (2). We followed a non-linear curve
fitting method. Initially, the parameters ($\theta_{1}$, C$_{eff}$,
$\xi$ and $\theta_{2}$) were allowed to take initial values and
iterated 10 times. As soon as the fitted curve comes close to the
experimental curve, we start to restrict the parameters one by one.
Finally, best fit curve was obtained by allowing all parameters to
vary, except $\theta_{2}$ keeping constant. The experimental data in
the paramagnetic state of the samples fitted with equation (2) are
shown in Fig. 3. The fit on susceptibility data in the paramagnetic
regime according to equation (2) is excellent. The fit parameters
are shown in Table I. A comparative fits applying equation (1) and
(2) for mh98 and mh200 samples suggests that equation (1) may be
well valid at higher temperature, but equation (2) is more
appropriate to describe the magnetic behaviour over a wide
temperature range above T$_C$. We noted using equation (1) that the
paramagnetic Curie temperature ($\theta_{w}$) systematically
decreases as the material transforms from bulk polycrystalline phase
to nanocrystalline (NC) phase and then, to amorphous (AMP) phase.
The $\theta_{w}$ values remained positive for bulk as well as mh25,
mh61 and mh98 samples, where as $\theta_{w}$ becomes negative for
mh146 and mh200 samples. The negative value of $\theta_{w}$
indicates the introduction of antiferromagnetic exchange
interactions in the material as the particle size and crystalline
phase changed. As discussed in earlier report \cite{rnbJAP}, the
magnetic dynamics of the present material strongly depends on the
structural phase transformation, rather than the particle size
effects. The $\theta_{1}$ (obtained using equation (2)) also follows
the pattern of $\theta_{w}$, showing positive values only for mh25
and mh61 samples. The spin glass like feature in amorphous (mh146)
sample clearly proves the the reduction of ferromagnetic (FM)
exchange interactions or development of antiferromagnetic (AFM)
exchange interactions in nanocrystalline and amorphous samples,
because spin glass like feature needs sufficient amount of both
magnetic disorder and competition between FM/AFM interactions. On
the other hand, $\theta_{2}$ is always positive and change is not
drastic (within 9 K considering all milled samples). The positive
value of $\theta_2$ suggested the retaining of a strong double
exchange ferromagnetic interactions \cite{Anderson} both in
nanocrystalline and amorphous phase of
La$_{0.67}$Ca$_{0.33}$MnO$_{3}$ nanoparticles \cite{rnbJAP}. At the
same time, application of equation (2) suggests the ferrimagnetic
character of mechanical milled nanoparticle LCMN samples. Similar
magnetic behaviour was also observed in ferrimagnetic MnCr$_2$O$_4$
nanoparticles \cite{rnbMnCr2O4}. Some reports \cite{Huang,Li} also
attempted to explain the magnetization data in the paramagnetic
regime of ferromagnetic nanomaterials by following a simple
Curie-Weiss law (equation (1)), but those data seem to be more
appropriate to the ferrimagnetic
description (equation (2)).\\
The validity of ferrimagnetic equation (2) in our milled samples can
be examined by considering the core-shell spin structure of
nanoparticles, already proposed in earlier work \cite{rnbJAP}. The
existence of strong ferromagnetic order, even in the nanocrystalline
and amorphous phase, is essentially due to ferromagnetic ordered
core spins. On the other hand, magnetic disorder is confined mainly
in the shell part for nanocrystalline particles (NCR NP) and also
introduces in the core part for amorphous nanoparticles (AMP NP).
The shell spins may not be typical antiparallel with respect to
core, but effective spin moment of shell is obviously low in
comparison with ferromagnetic core and schematically shown in Fig.4.
Similar magnetic modulation was previously proposed for
antiferromagnetic nanoparticle \cite{CoRh2O4} and later applied for
ferromagnetic manganite nanoparticles \cite{Zhang}. This allows us
to consider the magnetic contributions form shell and core of a
nanoparticle equivalent to two unequal magnetic sublatticles (shown
in lower diagram of Fig. 4), as usually seen in a typical long
ranged ferrimagnet. It must be differentiated that two different
magnetic sublattices as we suggest here for the ferromagnetic
particles is not due to different crystal environments, i.e.,
tetrahedral and octahedral lattice sites of a typical ferrite
consisting of two magnetic sublattices of antiparallel directions
\cite{ferri,rnbMnCr2O4}. Based on the experimental observations, the
concept of two different magnetic structure could be a realistic
approach for describing the magnetic properties of ferromagnetic
nanomaterials. Recently, similar concept was modelled by C.R.H. Bahl
et al. \cite{Bahl} and M. Vasilakaki et al. \cite{Vasil}.

\section{CONCLUSIONS}
La$_{0.67}$Ca$_{0.33}$MnO$_{3}$ ferromagnet exhibited many
interesting features in the nanocrystalline and amorphous phase, as
an effect of increasing disorder in core-shell spin morphology and
lattice structure. The present work clearly provides the evidence of
ferrimagnetic character in ferromagnetic
La$_{0.67}$Ca$_{0.33}$MnO$_{3}$ (LCMN) nanoparticles. The
ferrimagnetic concept, as propsed in this work, is interesting and
could be applied for the understanding of basic mechanism in many
ferromagnetic nanoparticles. Especially, this approach could be more
effective for the proper demonstration of the effect of core-shell
spin structure in ferromagnetic nanoparticles.\\

\textbf{Acknowledgement:} We thank A. Poddar of Saha Institute of
Nuclear Physics and CIF, Pondicherry University for magnetic
measurements.

\pagebreak
\begin{table*}
\caption{\label{tab:table1} The particle size (d) of the milled
samples are determined from the TEM data. The fit parameters (C and
$\theta_w$) were obtained using simple Curie-Weiss law (equation 1).
The parameters (C$_{eff}$, $\theta_1$, $\theta_2$ and $\xi$) were
obtained using equation (2) for different milled samples. It may be
mentioned that the present values of C and $\theta_w$ (K) using
temperature range 330 K to 400 K are slightly different from the
values C $\sim$ 0.0176 and $\theta_w$ $\sim$ 275 K using temperature
range 300 K to 340 K and previously repored \cite{rnbJAP}.}
\begin{ruledtabular}
\begin{tabular}{cccccccc}
Sample & d & C (K g Oe/emu) & $\theta_w$ (K) & C$_{eff}$ (K g
Oe/emu) & $\theta_1$ (K) & $\theta_2$(K) & $\xi$ (arb. unit)
\\\hline
Bulk & few $\mu$m & 0.196 & 270 & -- & -- & -- & -- \\
mh25 & 65 nm & 0.0275 & 200 & 0.035(3)  & 120$\pm$4 & 251 & 106450 $\pm$2200 \\
mh61 & 12 nm  & 0.0365 & 100 & 0.043(2) & 33$\pm$3 & 250 & 53400$\pm$520 \\
mh98 & 16 nm  & 0.0244 & 20 & 0.043(2) & -46$\pm$4 & 247 & 91800$\pm$1000 \\
mh146 & 60 nm & 0.0251 & -80 & 0.024(6) & -87$\pm$2 & 243 & 80000$\pm$1200 \\
mh200 & 90 nm & 0.0245 & -97 & 0.024(3) & -106$\pm$3 & 242 & 62000$\pm$800 \\

\end{tabular}
\end{ruledtabular}
\end{table*}


\begin{thebibliography}{99}
\bibitem{Ramirez} A.P. Ramirez, S. W. Cheong, and P. Schiffer,
J. Appl. Phys. \textbf{81}, 5337 (1997).
\bibitem{Gubin} S.P. Gubin, Y.A. Koksharov, G.B. Khomutov and G. Y. Yurkov, Russ.
Chem. Rev. \textbf{74}, 489 (2005).
\bibitem{Battle} O. Iglesias, A.Labarta, and X. Batlle, J. Nanoscience and Nanotech.
\textbf{8}, 2761 (2008).
\bibitem{CoRh2O4} R.N. Bhowmik,R. Nagarajan, and R. Ranganathan,Phys. Rev.B \textbf{69}, 054430 (2004).
\bibitem{Bahl} C.R.H. Bahl, J. Garde1,, K. Lefmann, T.B.S. Jensen, P.-A. Lindg, D.E. Madsen,
and S. Mørup, Eur. Phys. J. B \textbf{62}, 53 (2008).
\bibitem{Vasil} M. Vasilakaki and K. N. Trohidou, Phys. Rev. B \textbf{79}, 144402
(2009).
\bibitem{Muri} M. Muroi, P.G. McCormick and R. Street, Rev. Adv.
Matter. Sci. \textbf{5}, 76 (2003).
\bibitem{Masala} O. Masala and R. Seshadri, J. Am. Chem. Soc. \textbf{127}, 9354
(2005).
\bibitem{Bouz} G. Bouzerar, R. Bouzerar, and O. Cépas, Phys. Rev. B \textbf{76},144419
(2007).
\bibitem{Coral} V. Corral-Flores, D. Bueno-Baques, D. Carrillo-Flores, and J.A. Matutes-Aquino,
J. Appl. Phys. \textbf{99}, 08J503 (2006).
\bibitem{rnbJAP} R. N. Bhowmik, A.Poddar, R. Ranganathan, and Chandan Mazumdar, J. Appl. Phys.
\textbf{105}, 113909 (2009).
\bibitem{ferri} J. S. Smart, Am. J. Phys. \textbf{23}, 356 (1955).
\bibitem{rnbMnCr2O4} R.N. Bhowmik, R. Ranganathan, and R. Nagarajan, Phys. Rev. B \textbf{73},
144413(2006).
\bibitem{Anderson} P.W. Anderson and H. Hasegawa Phys. Rev. \textbf{100}, 675
(1955).
\bibitem{Huang}Y. H. Huang, J. Linde´n, H. Yamauchi, and M.Karppinen,Chem. Mater. \textbf{16}, 4337 (2004).
\bibitem{Li} X.H. Li, Y.P. Sun, W.J. Lu, R. Ang, S.B. Zhang, X.B. Zhu, W.H. Song, and J.M. Dai
Solid State Comm. \textbf{145}, 98 (2008).
\bibitem{Zhang} T. Zhang, T.F. Zhou, T. Quian, and X.G. Li, Phys. Rev. B \textbf{76}, 174415
(2007).

\end{thebibliography}
\end{document}